\documentclass[%
 reprint,
 amsmath,amssymb,
 aps,nofootinbib,
]{revtex4-2}

\usepackage{graphicx}
\usepackage{dcolumn}
\usepackage{bm}
\usepackage{xcolor}
\usepackage{soul}
\usepackage{comment} 

\usepackage{hyperref}

\begin{document}


\title{Can neutrino-assisted early dark energy models ameliorate the $H_0$ tension in a natural way?}


\author{Diogo H. F. de Souza}
\email{diogo.henrique@unesp.br}

\author{Rogerio Rosenfeld}
 \email{rogerio.rosenfeld@unesp.br}
\affiliation{
 Instituto de Física Teórica da Universidade Estadual Paulista and ICTP South American Institute for Fundamental Research, \\ R. Dr. Bento Teobaldo Ferraz, 271, Bloco II, Barra-Funda - São Paulo/SP, Brasil.
}

\date{\today}

\begin{abstract}
The idea of neutrino-assisted early dark energy ($\nu$EDE), where a coupling between neutrinos and the scalar field that models early dark energy (EDE) is considered, was introduced with the aim of reducing some of the fine-tuning and coincidence problems that appear in usual EDE models. 
In order to be relevant in ameliorating the $H_0$ tension, the contribution of EDE to the total energy density ($f_\text{EDE}$) should be around 10\% near the redshift of matter-radiation equality.
We verify under which conditions $\nu$EDE models  can fulfill these requirements for a model with a quartic self-coupling of the EDE field
and an exponential coupling to neutrinos. 
We find that in the situation where the EDE field is frozen initially, the contribution to $f_\text{EDE}$ can be significant but it is not sensitive 
to the neutrino-EDE coupling and does not address the EDE coincidence problem. On the other hand, if the EDE field starts already dynamical at the minimum of the effective potential, it tracks this time-dependent minimum that presents a feature triggered by the neutrino transition from relativistic to nonrelativistic particles. This feature generates $f_\text{EDE}$ in a natural way at around this transition epoch, that roughly coincides with the matter-radiation equality redshift. 
For the set of parameters that we considered we did not find values that satisfy the requirements on the background cosmological evolution to mitigate the Hubble tension in a natural way in this particular $\nu$EDE model.

\end{abstract}


\maketitle

\section{Introduction}
\label{sec:Introduction}

Despite of its great success in being able to describe a vast amount of observables from different
epochs of the evolution of the Universe, the standard spatially flat $\Lambda$CDM model
has recently being under stress mainly from the so-called Hubble tension arising from measurements of the Hubble constant from the cosmic microwave background (CMB) data in comparison to local measurements (for recent reviews, see \cite{di2021realm,kamionkowski2022hubble,Schoeneberg_2021_H0}).
One of the most promising solutions to the Hubble tension is a class of models called Early Dark Energy (EDE) \citep{de_early_times, fluidEDE, rocknroll, ede_can_restore,smith2020oscillating,Herold:2022iib}.
The current status of EDE models proposed to resolve the Hubble tension was recently reviewed in \cite{Poulin:2023lkg}.

A non-negligible amount of a new dark energy component that becomes dynamical at around the matter-radiation equality era and quickly dissipates afterwards would decrease the sound horizon that enters the indirect measurement of the Hubble constant from the CMB data, leading to an increase in the estimated value of the Hubble parameter today; hence EDE can ameliorate the Hubble tension \citep{edeshiftshorizon}.  In fact, the existence of EDE has already been proposed in the context of late quintessence models with global attractor solutions and its effect on the CMB and structure formation were studied in \citep{caldwell2003early} and \citep{bartelmann2006non}. 
In the more recent versions of EDE motivated by the Hubble tension, a scalar field with an axion-like or a pure $\phi^{4}$ potential was proposed  to model the EDE component \cite{fluidEDE,Smith_2020,rocknroll}. It has been shown that if EDE contributes with a fraction $f_{\text{EDE}} \approx  10\%$ of the total energy density of the Universe in an epoch with redshift around $\log_{10}(z)\sim3.5$, its presence is expected to decrease the statistical significance of the Hubble tension (see, eg, \cite{ede_can_restore}).

However, in the standard EDE scenario, some level of fine-tuning is necessary in order to solve the Hubble tension: the scalar field mass and its initial conditions need to have specific values in order for the EDE model to contribute the required amount of energy density at around the matter-radiation equality epoch. This is sometimes referred to as the EDE coincidence problem.

Some models were proposed to address the EDE coincidence problem by introducing a coupling of the EDE field to matter fields.
For instance, a coupling of EDE to dark matter as a trigger for its dynamics was recently studied in \cite{karwal2022chameleon,Lin:2022phm}.

Another similar mechanism was proposed in \cite{Sakstein2019EarlyDE,CarrilloGonzlez2020NeutrinoassistedED}, where a coupling of the EDE field to 
neutrinos was introduced.
There are two distinct consequences of introducing such a coupling: (i) a modification in the original EDE field potential leading to a new effective potential 
and (ii)
a dynamical effect of a ``kick'' to the EDE field. This kick is 
associated with the transition of neutrinos from being relativistic to nonrelativistic particles when the temperature of the universe falls below $T = { \cal O}(m_\nu)$, which causes a jump in the trace of the neutrino energy-momentum tensor.
This jump can displace the value of the EDE field that then starts its oscillations and dissipation. 
This mechanism can naturally trigger the EDE field at around the matter-radiation equality since the temperature at this epoch is approximately at the scale of neutrino masses. 

In this paper, we explore this neutrino-assisted EDE model ($\nu$EDE) in more detail.
In order to do this, we developed an implementation of the $\nu$EDE model in \texttt{CAMB}\footnote{https://github.com/cmbant/CAMB} at the background level to analyse the evolution of the EDE scalar field in the time-dependent effective potential 
given some initial conditions and model parameters and the computations are consistently performed in the Einstein frame.
We study two cases for the initial conditions: the field starting either at some arbitrary value or starting at the 
the minimum of the effective potential (see, eg, Brax et al. \cite{brax2004detecting}).
In the first case we consider the field initially frozen at an arbitrary value by Hubble friction ($|V''_{\text{eff}} / H^2| <1$). 
In the second case, we chose parameters of the model such that the field has already relaxed to the minimum of the effective potential.

We show that in the first case the model can result in a cosmologically significant amount of $f_{\text{EDE}}$ at the redshift
required to ameliorate the Hubble tension. However, using the  approach proposed in \cite{Lin:2022phm}, we point out that this result 
is not due to the neutrino-EDE coupling and it does not address the EDE coincidence problem.
In the second case, the model provides a natural value for the initial condition of the EDE field given by the minimum of the effective potential \cite{Sakstein2019EarlyDE, CarrilloGonzlez2020NeutrinoassistedED,CarrilloGonzalez:2023lma}.
Nevertheless, we find that the contribution to $f_{\text{EDE}}$ is too small for addressing the Hubble tension.

An important feature of the $\nu$EDE model is the neutrino kick \cite{Sakstein2019EarlyDE,CarrilloGonzlez2020NeutrinoassistedED} which may occur 
by a similar mechanism previously studied in \cite{brax2004detecting,Coc:2006rt,Erickcek:2013oma,Erickcek:2013dea}. 
The latter are chameleon models addressing the late dark energy with a different class of self-interaction potentials.
In the model studied here we do not see the effects of the neutrino kick on the EDE field with a displacement from 
the minimum of the effective potential by solving its field evolution equation.
Rather, we show that the jump in the trace of the neutrino energy-momentum tensor
is reflected in a change in the minimum of the effective potential that the EDE is following adiabatically.
It is this change that leads to an increase in $f_{\text{EDE}}$.

This paper is organized as follows. In Section \ref{sec:ModelDynamics} we present a summary of the model, and proceed to Section \ref{sec:finetuning} for a description of the methods used in the background analysis. Our results are presented in Section \ref{sec:initial_value} where we differentiate the discussion in the two regimes described above: in Section \ref{sec:phi_initially_frozen} the field is initially frozen by Hubble drag while in Section \ref{sec:phi_initially_dynamical} the field is initially dynamical. In both scenarios we compute the fraction of EDE and assess the importance of the neutrino-EDE coupling.  We present our conclusions in Section \ref{sec:conclusion}.

\section{Neutrino-assisted EDE}
\label{sec:ModelDynamics}

The idea of neutrinos coupled to a dark energy field was introduced in the so-called ``Mass-Variable Neutrinos" (MaVaN) models, since this coupling induces a variation of the neutrino mass due to the cosmic evolution of the background dark energy (also known as quintessence)  field \cite{fardon2004dark,peccei2005neutrino}.
 The effects of mass-varying neutrinos on the CMB anisotropies and large scale structures (LSS), both at the background and perturbative levels, were studied in \cite{Brookfield:2005td, brookfield2006cosmology,ichiki2008primordial}. 
The issue of fine-tuning was discussed in the context of quintessence models in \cite{Franca:2002iju} and more recently 
in MaVaN models in  \cite{maziashvili2023avoiding}. 
 We now present a brief summary of the model.
 
Consider a general action of the following form:
\begin{align}\label{totalaction}
    S &=\int d^4x\sqrt{-g}\left[\frac{1}{2}M_{pl}^2R-\frac{1}{2}\nabla_\mu\phi\nabla^\mu\phi-V(\phi)\right]\nonumber\\
    &+ S_{dm}[\Psi_{dm},\Tilde{g}^{dm}_{\mu\nu}] + S_{b}[\Psi_{b},\tilde{g}^b_{\mu\nu}]\nonumber\\
    &+ S_\gamma[\Psi_\gamma,\tilde{g}^\gamma_{\mu\nu}] +  S_\nu[\Psi_\nu,\Tilde{g}^\nu_{\mu\nu}],
\end{align}
where $\Psi_{dm}$, $\Psi_{b}$, $\Psi_\gamma$ and $\Psi_\nu$ are the dark matter, baryons, photon, and neutrino fields, respectively, $M_{Pl}\equiv(8\pi G)^{-1/2}$ is the reduced Planck mass and $g$ is the determinant of the metric $g_{\mu\nu}$ in the Einstein-frame. 
The Jordan-frame metric, $\tilde{g}_{\mu\nu}$, for each species are related with $g_{\mu\nu}$ by $\tilde{g}^i_{\mu\nu}=A^2_i(\phi)g_{\mu\nu}$ where $i$ represents all the components cited above. It is interesting to note that in the case $A_{dm}=A_b=A_\gamma=A_\nu=1$ and $V(\phi) = \mu^4[1-\text{cos}(\phi/f)]^n$, where $f$ is the energy scale where a shift symmetry is spontaneously broken and  $\mu$ is a small 
explicit shift symmetry-breaking breaking energy scale that generates an axion-like potential, one 
obtains the canonical uncoupled axion-like EDE model \cite{de_early_times, fluidEDE}. 
When $A_b=A_\gamma=A_\nu=1$, $A_{dm}\equiv A=\text{exp}(\beta\phi/M_{pl})$ and $V(\phi)=\lambda\phi^4$ we obtain the chameleon EDE model, recently studied in \cite{karwal2022chameleon}. 
Finally, when $A_{dm}=A_b=A_\gamma=1$, $A_\nu\equiv A=\text{exp}(\beta\phi/M_{pl})$ and $V(\phi)=\lambda\phi^4/4$ we obtain the neutrino-assisted EDE ($\nu$EDE) model studied in \cite{Sakstein2019EarlyDE,CarrilloGonzlez2020NeutrinoassistedED}. This is the model considered in the following. 
Notice that $\nu$EDE has the same framework and is described by the same equations as for MaVaNs model, the main difference being that for the latter case, the quintessence potential is $V(\phi)=V_0\exp(-\sigma\phi)$ which is required to achieve the late time cosmic acceleration.  

The coupling function that changes the metric between the Einstein and Jordan frames can be related to the time-varying neutrino mass as $m_\nu(\phi)=m_{\nu0}A(\phi)$ where $m_{\nu0}$ is the neutrino mass today. 
The average neutrino energy density ($\rho_\nu$) and pressure ($p_\nu$) are given in terms of an integral over the comoving momentum ($q$) of the usual unperturbed Fermi-Dirac distribution $f_0(q)$:
\begin{align}
    \rho_\nu(\phi) &= \frac{1}{a^4}\int q^2dqd\Omega\epsilon f_0(q),\label{nuEnergyDensity}\\
    p_\nu(\phi) &=\frac{1}{3a^4}\int q^2dqd\Omega \frac{q^2}{\epsilon}f_0(q),\label{nuPressure}\\
    f_0(q) &= \frac{g_\nu}{h^3_P}\frac{1}{e^{q/{k_BT_{\nu0}}}+1},\label{unperFermiDirac}\\
    \epsilon^2 & = q^2 + a^2m_\nu^2(\phi),\label{modfied_energy}
\end{align}
where $g_\nu$, $T_{\nu0}$, $h_P$, $k_B$ are the number of spin degrees of freedom, the neutrino temperature today, the Planck and Boltzmann constant, respectively.  In equation \eqref{unperFermiDirac} we have assumed that neutrinos decoupled from the primordial plasma when they are still relativistic, so the phase-space distribution preserve the Fermi-Dirac form with a dependence only on the comoving momentum.

Taking the time derivative of Eq.~\eqref{nuEnergyDensity} and using that the energy conservation is determined by the coupled system of neutrinos and the scalar field, we obtain the modified Klein-Gordon equation
\begin{align}\label{modfiedKleinGordon}
    \ddot{\phi}+3H\dot{\phi} +\frac{dV(\phi)}{d\phi}&= -\frac{d\,\text{ln}\,m_\nu(\phi)}{d\phi}\left(\rho_\nu-3p_\nu\right),
\end{align}
where dot is the derivative with respect to time. The self-interaction EDE potential assumed in this work is
\begin{equation}
V(\phi)=\lambda\phi^4/4. 
\end{equation}
This is the potential used in the so-called ``rock 'n' roll" models of EDE \cite{Agrawal:2019lmo}. 

In addition, it is interesting to notice that taking the derivative of Eq.~\eqref{nuEnergyDensity} with respect to the scalar field results in
\begin{equation}
d\rho_\nu(\phi)/d\phi=(d\,\text{ln}\,m_\nu/d\phi)(\rho_\nu-3p_\nu). 
\label{derivative_rho}
\end{equation}

For definiteness we are going to adopt a coupling of the form:
\begin{equation}\label{conformalCoupling}
    A(\phi) = \exp(\beta\phi/M_{pl}),
\end{equation}
where $\beta$ is a dimensionless constant factor giving the coupling strength between neutrinos and the EDE field.
The fact that in EDE models the energy density must dissipate quickly after recombination ensures that $m_\nu(\phi) \rightarrow m_{\nu0}$ today.
With this choice of interaction the Klein-Gordon equation takes the familiar form:
\begin{equation}\label{modfiedKleinGordon2}
    \ddot{\phi}+3H\dot{\phi}+\frac{dV(\phi)}{d\phi} = -\frac{\beta}{M_{pl}}\Theta_{\nu},
\end{equation}
where $\Theta_{\nu}\equiv-\rho_\nu+3p_\nu$ is the trace of the energy-momentum tensor for the neutrinos.
Therefore, one can define an effective potential where
\begin{equation}\label{well_defined_derivative}
    V'_\text{eff} = V'(\phi) - \frac{\beta}{M_{pl}}\Theta_{\nu}.
\end{equation}
and using equation (\ref{derivative_rho}) we write the redshift-dependent effective potential as
\begin{equation}
    V_\text{eff}(\phi) = V(\phi) + \rho_\nu(\phi,z).
    \label{eq:Veff}
\end{equation}

An important quantity throughout this work is the redshift-dependent minimum of the effective potential given by \cite{CarrilloGonzlez2020NeutrinoassistedED}\footnote{Notice that since we are working in the Einstein frame there is an implicit dependence of  $\Theta_\nu$ on the field and the coupling constant $\beta$. This dependence can be neglected only if $|\beta\phi/M_{Pl}|\ll1$. We take this dependence into account in our numerical code.}:
\begin{align}
    \phi_\text{min} &= -\left(\frac{\beta\Theta_\nu}{\lambda M_{Pl}}\right)^{1/3},
    \label{eq:phimin}
\end{align}

The free parameters of the model are the scalar field self-coupling constant $\lambda$, a parameter $\beta$ that characterizes the scalar field coupling to neutrinos, 
and the initial conditions for the field $\phi_i$ and $\dot{\phi}_i$. In the following we will adopt three degenerate neutrinos with $\sum m_{\nu0}=0.3$ eV which 
is below the upper bound found in \cite{DiValentino:2019dzu} with \textit{Planck} data only.

\section{A background analysis of the neutrino-assisted EDE}
\label{sec:finetuning}
As previously stated, the free parameters of the model are $\lambda$, $\beta$, and $\phi_i$ (we will assume $\dot{\phi}_i=0$, i.e. the field is frozen initially). 
In order to gain some intuition on the parameter $\lambda$, notice that in the axion-like EDE potential with $V(\phi) = \mu^4[1-\text{cos}(\phi/f)]^n$, an expansion around the minimum for $n=1$ gives the axion mass $m_a = \mu^2/f$  and the quartic coupling $\lambda=m_a^2/f^2$.
In scalar field models, it is well-known that the field becomes dynamical (i.e. it ``thaws") when its mass is of the order of the Hubble parameter.
Therefore, if we want the EDE field to become dynamical before recombination ($H \approx (\text{1 eV})^2/M_{Pl} \approx {\cal O}( 10^{-27})$ eV) and assuming a typical symmetry breaking scale of 
$f = 10^{15}$ GeV one expects very small values of the self-coupling of $\lambda = {\cal O} (10^{-100})$.  Even in anharmonic potentials with $n>1$, this behavior is present \cite{fluidEDE}, and in our case the value of 
$\lambda$ controls the epoch when EDE becomes dynamical, with a redshift usually denoted by  $z_c$.
More generally, the field becomes dynamical when \cite{brax2004detecting} 
\begin{equation}\label{field_is_dynamical}
    \frac{V''_\text{eff}}{H^2}=\mathcal{O}(1).
\end{equation}

Another important quantity in EDE models is the maximum contribution of the EDE component to the total energy budget of the universe at $z_c$, denoted by 
$f_\text{EDE}$:
\begin{equation}
    f_\text{EDE}(z_c) = \frac{\rho_\text{EDE}(z_c)}{\rho_\text{tot}(z_c)},
\end{equation}
and it is determined by the model parameters. In fact, in an EDE model there is a simple relation between $\lambda$, $\phi_i$ and $z_c$:
\begin{equation}
    \lambda \approx \frac{T_0^4}{9 M_{Pl}^2 \phi_i^2 } (1+z_c)^4,
    \label{eq:unfrozen}
\end{equation}
where $T_0$ is the CMB temperature today.



The solution to the Hubble tension in EDE models should fulfill the basic requirements \cite{Herold:2022iib,Lin:2022phm}
\begin{align}\label{Lin_requirements}
    f_\text{EDE}\sim0.1~\text{and}~\log_{10}z_c\sim3.5
\end{align}
and that the EDE field is dynamical just prior to matter-radiation equality.
However, usual EDE models suffer from a coincidence problem due to the arbitrary initial conditions.
Extensions of EDE models including interactions with other components such as dark matter \cite{karwal2022chameleon} or neutrinos \cite{Sakstein2019EarlyDE,CarrilloGonzlez2020NeutrinoassistedED} may address the EDE coincidence problem via some triggering mechanism \cite{Lin:2022phm}. 
In order to investigate the impact of the new coupling $\beta$ to the fraction of EDE produced in the extended model we introduce
the quantity 
\begin{align}\label{nu_triggered_decay}
    \Delta\equiv\frac{f_\text{EDE}}{f_\text{EDE}^{\beta=0}}, 
\end{align}
which is defined by the ratio between the fraction of EDE for the coupled ($\beta\neq0$) and uncoupled ($\beta=0$) scenarios. In the limit that $\Delta\rightarrow1$ the $f_\text{EDE}$ is mainly sourced by the bare potential \cite{Lin:2022phm}.

\begin{figure}[!htp]
    \centering
    \includegraphics[scale=.5]{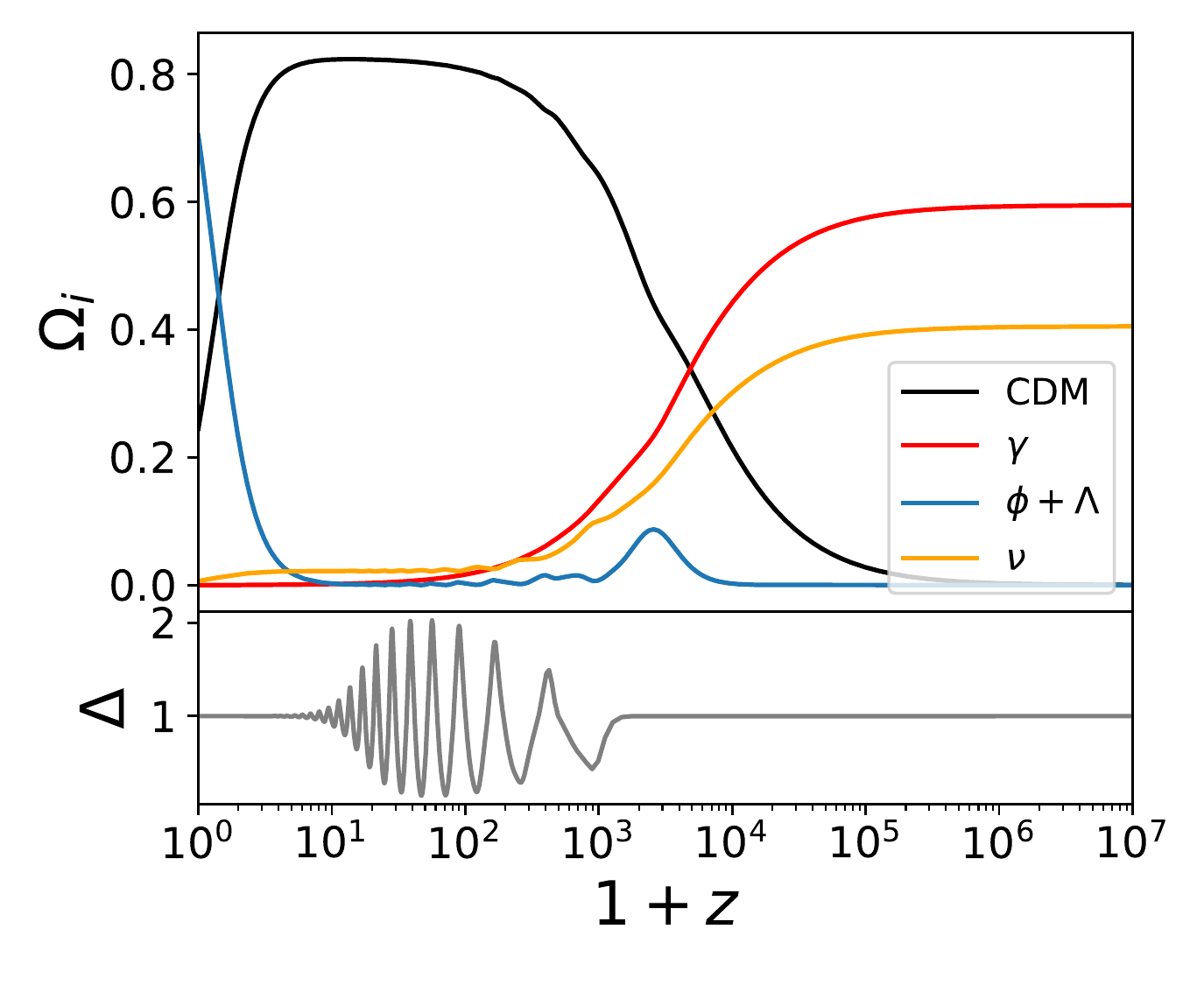}
    \caption{The background evolution for cold dark matter (CDM), photons ($\gamma$), neutrinos $\nu$ and the scalar field plus a cosmological constant ($\phi+\Lambda$). The parameters used are $\beta=10$, $\phi_i=-0.6\,{M}_{Pl}$ and $\log_{10}\lambda=-109$. The surge of EDE is around at $z\sim2\times10^3$. The bottom panel shows the quantity defined in Eq.~\eqref{nu_triggered_decay} which is the ratio between the coupled and uncoupled cases. 
    }
    \label{fig:BackgroundEvolution}
\end{figure}

In the top panel of Fig.~\ref{fig:BackgroundEvolution} we show an example
of the background evolution for cold dark matter (CDM), photons ($\gamma$), neutrinos $\nu$ and the EDE scalar field plus a cosmological constant ($\phi+\Lambda$) where the conditions in Eq.~\eqref{Lin_requirements} are satisfied. In the bottom panel one can see 
that in this case the quantity $\Delta$ is close to 1 at the EDE peak ($z \sim 2 \times 10^3$), indicating that the coupling to neutrinos is not relevant to generating $f_\text{EDE}$.
We will show below cases with the opposite situation.


\section{Initial value of the EDE field}\label{sec:initial_value}
Previously we assumed an arbitrary initial value for the EDE field.
We now want to address this issue in more detail. We identify two situations: 
a) the field is still frozen initially and starts to follow the minimum of the 
effective potential after its thawing
and b) the field is already dynamical initially and therefore it follows the minimum of the effective potential.
This separation is somewhat arbitrary since at large enough redshifts the field is frozen.
However, in practical terms, we need to choose an initial redshift from which we start evolving the field numerically.
The separation between these two regimes at a given redshift $z_i$ can be estimated by the condition
\begin{equation}\label{condition_for_dynamics}
   \frac{V''_\text{eff}(\phi_\text{min})}{H^2} = 3\lambda^{1/3}\beta^{2/3} \left(\frac{\Theta_\nu}{H^3M_{Pl}}\right)^{2/3}  \Bigg|_{z_i} = {\cal O} (1).
\end{equation}

\begin{figure}[!htp]
    \centering
    \includegraphics[scale=.5]{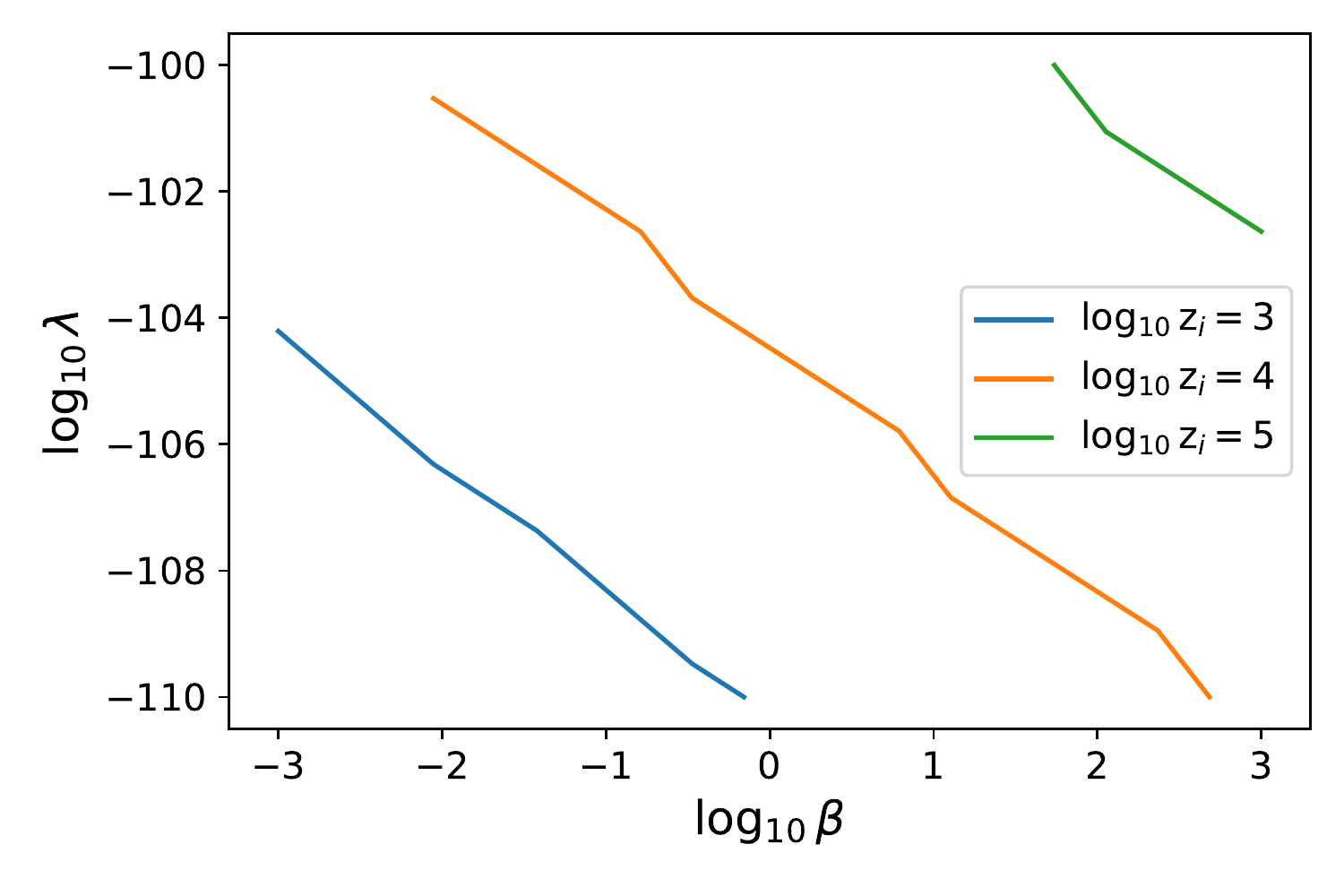}
    \caption{Values of $\lambda$ and $\beta$ for which $|V''_\text{eff}/H^2-1| = 0.1$ for $z_i = 10^3, 10^4$ and $10^5$. For regions above the curves the field is dynamical at the corresponding redshift.
    }
    \label{constrain_mphi2_over_H2}
\end{figure}

In Fig.~\ref{constrain_mphi2_over_H2} we show the contours in the parameter space $(\lambda,\beta)$
where the field is already dynamical at some given redshifts. The figure provides estimates of the values
of these coupling separating the two cases. 

The motivation for studying these two cases separately below is
that in the first case one could argue that the initial value of the frozen field 
does not have a natural value whereas in the
second case the value of the field is naturally close to the minimum of the effective potential.
Indeed we find different dynamical behaviours in these two cases.

\subsection{Field initially frozen}\label{sec:phi_initially_frozen} 
\begin{figure}[!htp]
    \centering
    \includegraphics[scale=.5]{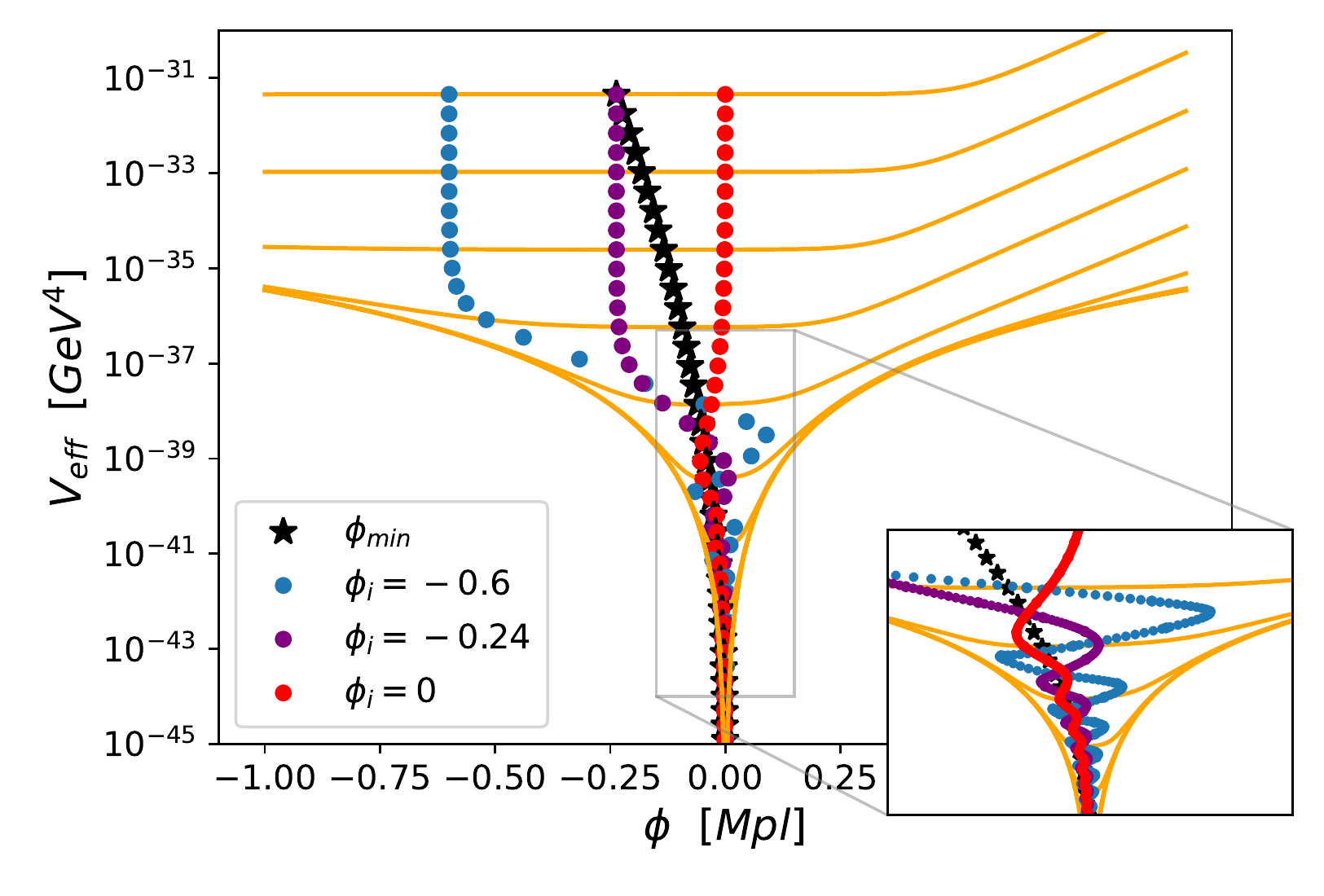}
    \caption{The orange lines are snapshots for the effective potential for decreasing values of redshift (the first line from top to bottom starts at $z_i=10^5$ and the redshift interval between two subsequent lines is $\log_{10}(z_n/z_{n+1})\approx0.41$). The black stars indicate the minimum of the effective potential and the blue, purple and red dots are the scalar field evolution starting at different initial conditions as indicated in the label. Notice that in the case represented by the purple dots the field starts at the minimum of the effective potential but does not follow it until thawing.
    Also is shown a zoom of the region where the field is dynamical and oscillates around the minimum of the effective potential.
    In this figure we used $\beta=10$ and $\lambda=10^{-109}$.}
    \label{effective_potential}
\end{figure}

In Figure~\ref{effective_potential} we plot the effective potential given in Eq.~(\ref{eq:Veff}) 
as a function of the field for different values of the redshift starting at $z_i = 10^5$ for $\beta=10$ and $\lambda = 10^{-109}$,
where $\rho_\nu(\phi, z)$ is given by Eq.~(\ref{nuEnergyDensity}). 
At high redshifts, the neutrino contribution dominates and it is a smooth function of $\phi$. At lower redshifts, the contribution
from $V(\phi)$ to the effective potential starts to become more relevant. 
For these parameters, we plot the evolution of the field for three different values of $\phi_i$: one starting at the minimum of the effective potential and the other two are chosen arbitrarily for comparison. We also show the minimum of the effective potential given by Eq.~(\ref{eq:phimin}) that, due to the coupling, evolves with redshift.


In all the cases the field is frozen at $z_i$, corresponding to the first line
in the plot. We find numerically that for the initial values of the field that we chose 
it remains frozen until approximately the redshift that satisfies Eq.~(\ref{eq:unfrozen}) is reached. 
Notice that as the neutrino contribution becomes small at lower redshifts, 
the minimum of the effective potential approaches the minimum of the bare potential $V(\phi)$ at $\phi=0$.

\begin{figure}[!htp]
    \centering
    \includegraphics[scale=.5]{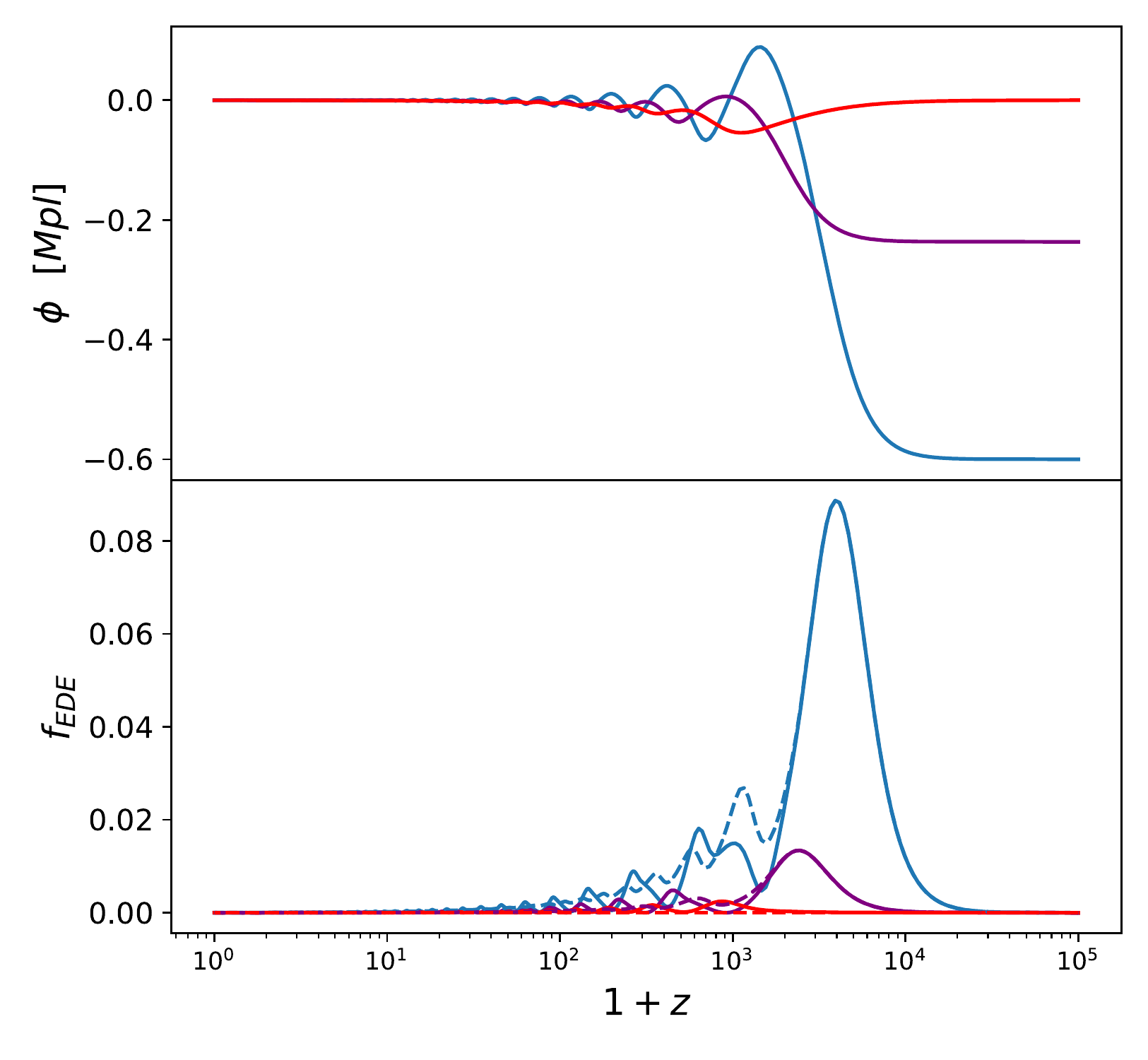}
    \caption{The scalar field evolution and the fraction of energy density for the same three initial conditions as in Fig.~\ref{effective_potential}: $\phi_i/M_{Pl}=-0.6, -0.24, 0$ for the blue, purple and red lines, respectively. The dashed lines in the bottom panel shows the uncoupled case with all other parameters kept fixed at the same values.}
    \label{field_fede_evolution}
\end{figure}

After thawing
the field starts to roll in the effective potential tracking and oscillating around its minimum. 
This behaviour of the field is shown both by the insert zoom in Figure \ref{effective_potential} and at the top panel of Figure \ref{field_fede_evolution}, where one can see the field relaxing to $\phi=0$, dissipating the energy density contained in it.

In the bottom panel of Figure \ref{field_fede_evolution} we show the contribution to $f_\text{EDE}$ in this example.
One can see that there could be a significant contribution depending on the initial value of the field. However, 
the coupling to neutrinos does not seem to play a large role in this case. 
We investigate this issue by computing the evolution of the field with the same initial conditions but without a coupling to neutrinos ($\beta=0$). This is shown as the dashed lines in the bottom panel of Figure \ref{field_fede_evolution}, where one can see that although the coupling introduces changes in the behaviour of the field the modification to the maximum of $f_\text{EDE}$ is negligible. In the case of the largest contribution, we find $|f_\text{EDE}/f_\text{EDE}^{\beta=0} -1| = 5 \times 10^{-4}$.
Therefore, in this case one needs the usual fine-tuning of initial conditions found in regular EDE models in order to have
a significant contribution to $f_\text{EDE}$.

\subsection{Field initially dynamical} \label{sec:phi_initially_dynamical}
In this subsection we are interested to study the case
where the field is already dynamical at the initial redshift considered.
The motivation is that in this case the field is already tracking the minimum of the effective
potential and hence its value is not arbitrary.
We will show below that in order for the field to become dynamical 
at an early redshift, a larger value of the self-coupling constant is necessary
but this leads to a small fraction of EDE.

\begin{figure}[!htp]
    \centering
    \includegraphics[scale=.53]{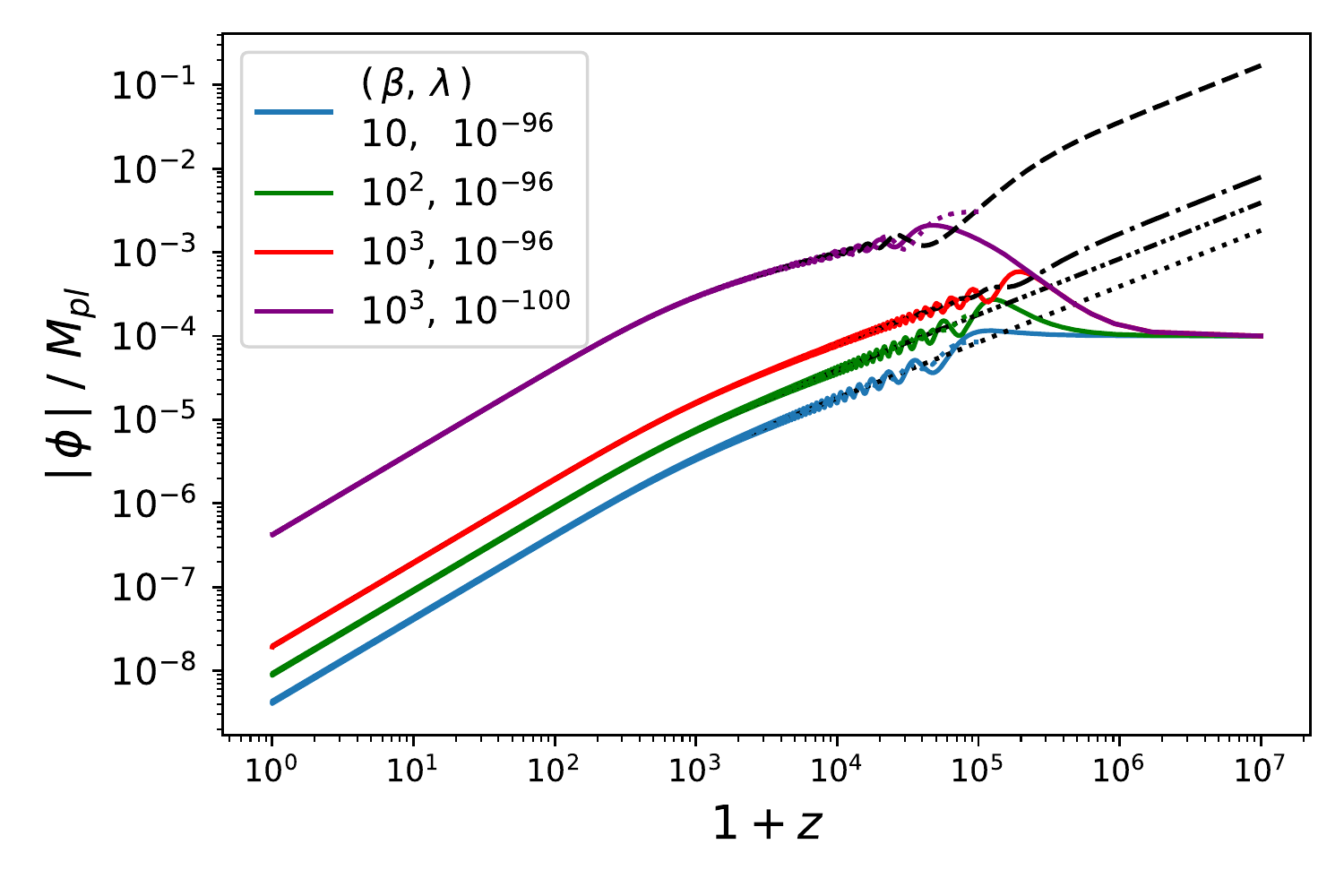}
    \caption{The full colored lines are the field evolution for $\phi_i=-10^{-4}M_{Pl}$ at $z_i=10^7$ and considering different values of the parameters $\beta$ and $\lambda$ as indicated in the label. Each field track its minimum as described by Eq.~\eqref{eq:phimin} and it is indicated by the different stylized black lines.  The dotted colored lines are the field evolution starting at $z_i=10^5$ in their respective minimum for the same set of parameters $\beta$ and $\lambda$.}
    \label{field_initially_dynamica}
\end{figure}

We begin by showing some examples of this behaviour in Figure \ref{field_initially_dynamica}.
We choose parameters in such a way that the field becomes 
dynamical and starts to oscillate around the minimum
of the effective potential at some $z = {\cal O}(10^5)$. 
More specifically, we start the field evolution at a high redshift $z = 10^7$ at $\phi_i/M_{Pl} = -10^{-4}$, where it is still frozen and show how it approaches the minimum of the effective potential at $z = {\cal O}(10^5)$ for different values of $\beta$ and $\lambda$. We also show that the difference is minimal when compared to the evolution starting at $z= 10^5$ with the field at the minimum of the effective potential.

\begin{figure}[!htp]
    \centering
    \includegraphics[scale=.45]{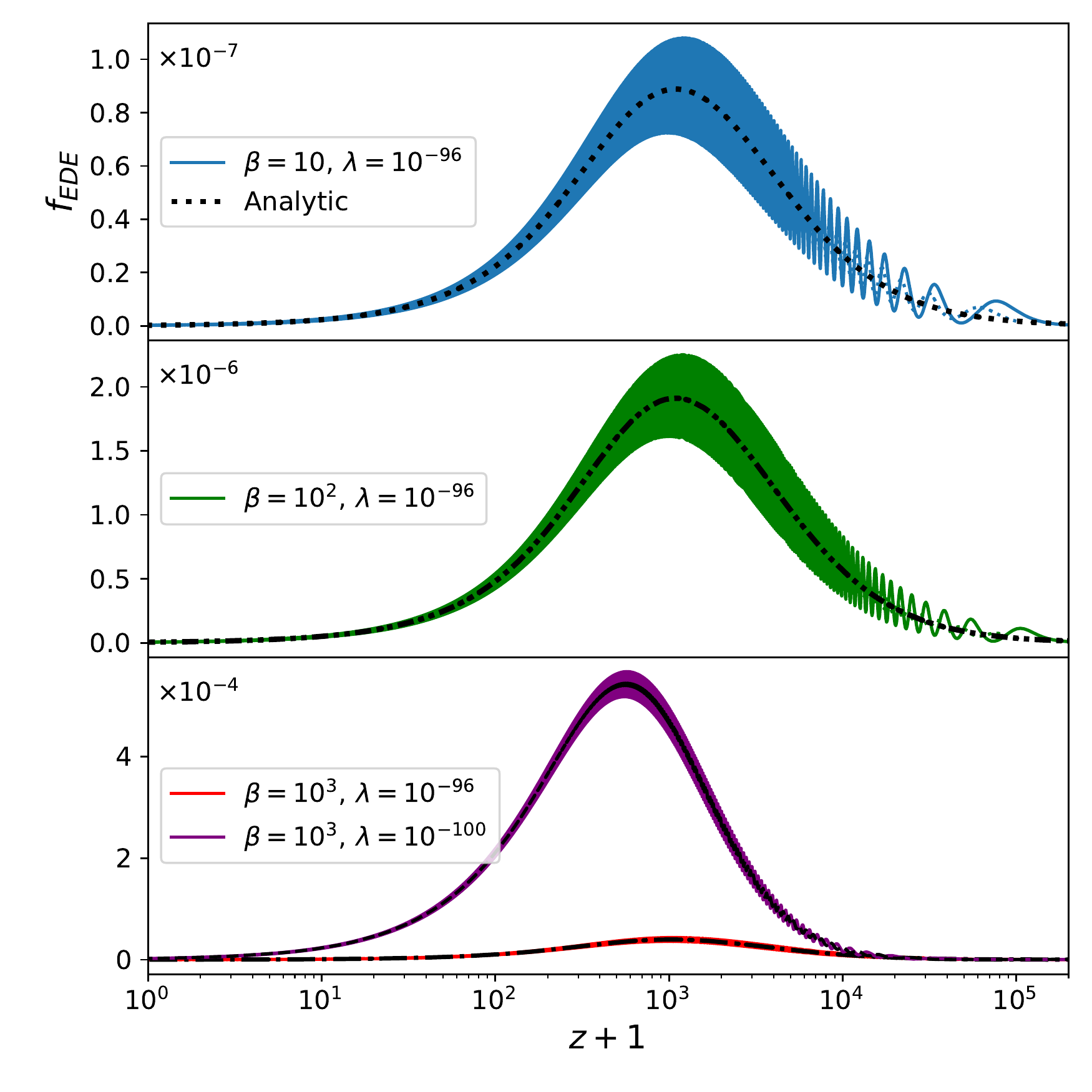}
    \caption{Fraction of EDE for the same set of parameters as in Fig.~\ref{field_initially_dynamica}. The different stylized black lines are the analytic prediction for the $f_\text{EDE}$ according to Eq.~\eqref{eq:fede}.}
    \label{fede_vsbeta_vs_lambda}
\end{figure}

We now turn to the contribution of the EDE field $\phi$ to the energy density of the Universe, showing in Figure \ref{fede_vsbeta_vs_lambda} the behaviour of $f_{\text{EDE}}$ for the cases
described above. 
We compare the numerical computation with the analytical estimate obtained by assuming that the EDE field is at the minimum of the effective potential (Eq.~\eqref{eq:phimin}) and that the dominant
contribution comes from the bare potential $V(\phi)$\cite{CarrilloGonzlez2020NeutrinoassistedED}:
\begin{equation}
    f_{\text{EDE}} =\frac{1}{12}\frac{\beta^{4/3}}{\lambda^{1/3}}\left(\frac{\Theta_\nu (\phi)}{H^{3/2}M_{Pl}^{5/2}}\right)^{4/3},
     \label{eq:fede}
\end{equation}
computed assuming that $\beta \phi/M_{Pl} <1$ in order to use the approximate expression Eq.~\eqref{eq:phimin}. 
However, in the equation above we use the value of $\phi$ from CAMB.
One can see that the analytical estimate is a reasonable approximation of the average of the oscillations in $f_{\text{EDE}}$.
We also notice that even for large values of the coupling ($\beta = 1000$) a cosmologically significant level of $f_{\text{EDE}}$ is not obtained\footnote{In principle it is possible to obtain a significant amount of EDE for still larger values of $\beta$. Indeed, we find that $\beta \geq {\cal O}(10^6)$ could in principle lead to $f_\text{EDE} = {\cal O}(10\%)$ but these large values of $\beta$ can not be considered natural.}. 
The same behaviour is also seen if one decreases the field self-coupling $\lambda$, as expected from Eq.~(\ref{eq:fede}) shown in the bottom panel of Fig.~\ref{fede_vsbeta_vs_lambda}: an increase in 
$f_{\text{EDE}}$ is obtained for smaller values of $\lambda$ at the cost of a decrease in the relevant redshift $z_c$.
A more thourough investigation including a comparison with data using a Bayesian analysis is necessary to reach a more definitive conclusion of whether this
case can help to ease the $H_0$ tension.

A prominent feature of $\nu$EDE models is the kick in the EDE field provided
by the coupling with neutrinos that occurs when neutrinos transition
from the relativistic to the nonrelativistic regimes. 
An intriguing possibility is that this mechanism could explain the EDE coincidence problem
since the neutrino mass energy scale roughly coincides
with the matter-radiation equality epoch.
We now want to assess the impact of the neutrino coupling in the
contribution of the EDE field to the total energy density.

\begin{figure}[!htp]
    \centering
    \includegraphics[scale=.5]{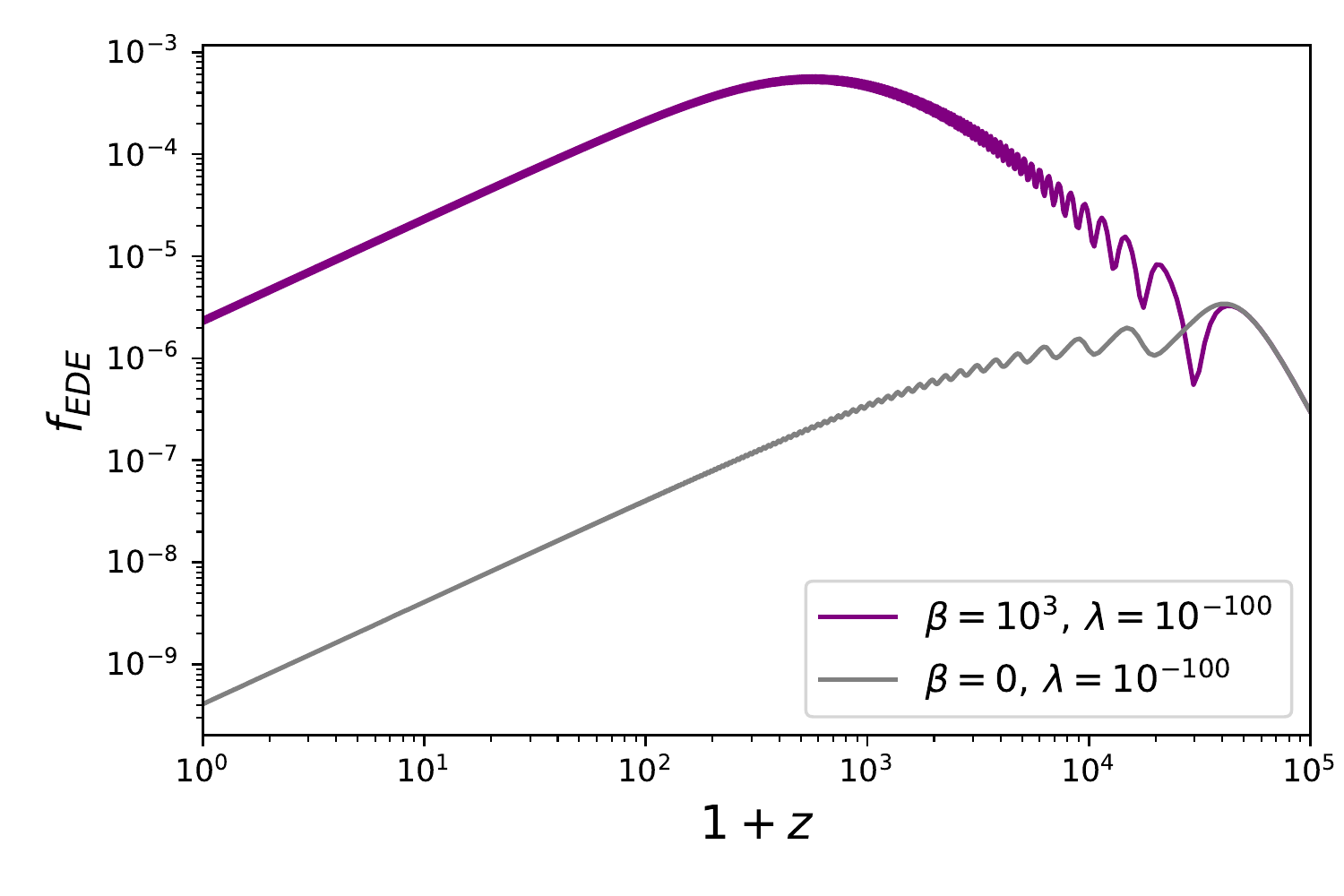}
    \caption{A comparison of $f_\text{EDE}$ between the coupled and uncoupled ($\beta=0$) scenarios with the field starting at the minimum of the effective potential at $z=10^5$ while keeping fixed all the other cosmological and model parameters. 
    The purple line is the case with largest $f_\text{EDE}$ shown in the bottom of Fig.~\ref{fede_vsbeta_vs_lambda}. 
    }
    \label{fede_coupled_uncoupled}
\end{figure}

In order to evaluate the impact of the coupling to neutrinos we 
will use the criteria in Eq.~\eqref{nu_triggered_decay} and compare the two cases, with and without coupling.
In both cases we start the field evolution at the minimum of the effective potential at $z=10^5$ in the situation where
the field is already dynamical
but in the first case we study the full model whereas in the second case we set the neutrino coupling $\beta$ to zero, 
keeping all other parameters equal.
In Figure \ref{fede_coupled_uncoupled} we show a comparison in $f_\text{EDE}$ for both cases, and one can see that the effect of the 
neutrino-EDE coupling is prominent. Although the contribution is the same initially, in the uncoupled case the field tends to zero (the minimum of the bare potential) while the coupled case follows Eq.~\eqref{eq:phimin} which for larger $\beta$ can significantly differ from zero.
    This results in a large difference between the two cases, which impacts $f_\text{EDE}$ by several orders of magnitude.

\section{Conclusion}
\label{sec:conclusion}
Neutrino-assisted early dark energy models were introduced with the aim of reducing some of the fine-tuning required in usual EDE models with respect to the parameters characterizing the scalar field self-coupling and its initial displacement. 
The idea that a ``neutrino kick" occurring around the matter-radiation equality epoch could trigger the EDE field is interesting for addressing the EDE coincidence problem.

In this work we developed an implementation of the $\nu$EDE model in \texttt{CAMB} at the background level to study $f_\text{EDE}$ and $\Delta$, quantifying the EDE scalar field contribution to the total energy density and the impact of the neutrino-EDE coupling, respectively.
We have performed this analysis considering two cases depending whether or not the EDE field is frozen at the initial redshift of its evolution. We find that if the field is initially frozen, its value can be arbitrary and 
the model can help in ameliorating the Hubble tension. However, since $\Delta\sim1$ the neutrino-EDE coupling is not relevant in this case and the EDE coincidence problem is not addressed. 

In contrast, as pointed out in \cite{CarrilloGonzlez2020NeutrinoassistedED,CarrilloGonzalez:2023lma}, the opposite scenario is attractive since there is a natural initial value for the field: the minimum of the effective potential, $\phi_\text{min}$, which is time-dependent. We show that the field follows $\phi_\text{min}$ without experiencing a significant dynamical kick from its evolution equation, ie without a displacement from the minimum. 
However, $\phi_\text{min}$ is affected by the change in the trace of the neutrino energy-momentum tensor, which generates 
a bump in $f_\text{EDE}$ since the field follows the minimum. 

We find that on average $f_\text{EDE}$ is well reproduced by the analytical prediction in Eq.~\eqref{eq:fede}. Furthermore, from Fig. \ref{fede_coupled_uncoupled}
one can see that in this regime the neutrino-EDE coupling plays a non-negligible role, leading to large modifications in $f_\text{EDE}$.
However, the attained value of $f_\text{EDE}$ does not satisfy the criteria in Eq.~(\ref{Lin_requirements}) for easing the Hubble tension. One might think that decreasing the value of the self-coupling $\lambda$
would increase the value of $f_\text{EDE}$, as follows from Eq.~\eqref{eq:fede}. This is correct but one should notice that this would also
imply in a field that is frozen until later times (since it results in a smaller $V''(\phi)$), which changes this model to the first case discussed above. In addition, we verified that couplings as large as $\beta=1000$ does not produce a significant amount of EDE. 

We conclude that $\nu$EDE models, with a self-interaction potential and a coupling of the type $V(\phi)\propto\phi^4$ and $A(\phi)=\exp(\beta\phi/M_{Pl})$, respectively, do not seem to naturally provide
a significant amount of early dark energy which is one of the criteria usually required for the model to  
ease the Hubble tension.
However, a more thourough investigation is certainly needed. 
Different potentials and couplings should be analysed, such as the ones proposed in Ref.~\cite{Lin:2022phm} where a model with EDE interacting with dark matter is studied using a monodromy-type potential and a $\phi^2$ coupling. 
Additionally, a successful model must not only be able to alleviate the Hubble tension at the background level but also produce
perturbations that agree with the current CMB and LSS data. Hence,
a comparison with data using Bayesian statistics at the perturbation level analysis (eg   \cite{brookfield2006cosmology,Brookfield:2005td,ichiki2008primordial}) is necessary in order to reach a more definitive conclusion of whether this class of models is capable of alleviating the $H_0$ tension and possibly the $S_8$ tension as well. 
It could be interesting to further study $\nu$EDE models using data from large galaxy surveys, such as the Dark Energy Survey and the imminent Rubin Telescope's Legacy Survey of Space and Time.

\begin{acknowledgments}
We are very much in debt to Mark Trodden, Jeremy Sakstein, Mariana González and Qiuyue Liang for their thoughtful criticism 
of the first version of this paper and for their patience in sharing their views on this subject. 
We would like to thank Vivian Miranda, David Mota, João Rebouças and especially Elisa Ferreira for useful suggestions about the text.
The work of DHFS is supported by a CAPES fellowship. RR is partially supported by a CNPq fellowship,
the project INCT do e-Universo, LIneA and Fapesp.
\end{acknowledgments}

\appendix

\newpage

\bibliography{nu-assisted}
\bibliographystyle{apsrev4-1}

\end{document}